\documentclass[aps,prl,longbibliography,twocolumn,superscriptaddress]{revtex4-2}
\usepackage{amsmath,amssymb}
\usepackage{graphicx}
\usepackage{subfigure}
\usepackage{verbatim}
\usepackage{amsfonts}
\usepackage{dcolumn}
\usepackage{bm}
\usepackage{color}
\usepackage{hyperref}
\hypersetup{
    colorlinks=true,
    citecolor= blue,
    linkcolor= blue,
    filecolor=blue,      
    pdfcreator = {\LaTeX\ and \flqq hyperref\frqq},
}
\usepackage[capitalize]{cleveref}
\usepackage{mathrsfs}
\usepackage{float}
\usepackage{multirow}
\usepackage[dvipsnames]{xcolor}

\newcommand{\beq}{\begin{eqnarray} }
\newcommand{\eeq}{\end{eqnarray} }
\newcommand{\Beq}{\begin{eqnarray*} }
\newcommand{\Eeq}{\end{eqnarray*} }
\newcommand{\Bmat}{\left(\begin{matrix}}
\newcommand{\Emat}{\end{matrix}\right)}
\newcommand{\up}{\uparrow}
\newcommand{\dn}{\downarrow}

\newcommand{\M}{\mathrm{M}}

\newcommand{\bq}{\mathbf{q}}

\newcommand{\bQ}{\mathbf{Q}}

\newcommand{\br}{\mathbf{r}}
\newcommand{\bR}{\mathbf{R}}
\newcommand{\ba}{\mathbf{a}}

\newcommand{\bdelta}{\bm{\delta}}
\newcommand{\scT}{\mathcal{T}}

\begin{document}

\title{Evidence for Deconfined Magnetic Order in the Kitaev-$J_3$ Model}

\author{Jiucai Wang}
\affiliation{School of Physics, Hangzhou Normal University, Hangzhou 311121, China}

\author{Chuan Chen}
\email{chenc@lzu.edu.cn}
\affiliation{School of Physical Science and Technology, Lanzhou University, Lanzhou 730000, China}
\affiliation{Lanzhou Center for Theoretical Physics, Key Laboratory of Quantum Theory and Applications of MoE,
Key Laboratory of Theoretical Physics of Gansu Province,
Gansu Provincial Research Center for Basic Disciplines of Quantum Physics, Lanzhou University, Lanzhou 730000, China}

\date{\today}

\begin{abstract}
We investigate the Kitaev-$J_3$ honeycomb model using variational Monte Carlo calculations combined with a vison-quasiparticle analysis of the parent Kitaev spin liquid (KSL). We provide evidence for deconfined magnetic phases in which zigzag or antiferromagnetic order coexists with remnant $\mathbb{Z}_2$ topological structure inherited from the KSL. The optimized variational wave functions retain multiple linearly independent topological sectors on a torus, whereas those of conventional ordered phases collapse to a single sector. The vison-quasiparticle analysis shows that magnetic order naturally arises from vison-pair condensation while single visons remain gapped, yielding a microscopic mechanism for magnetic ordering without immediate confinement. The resulting phases further host gapless spinons with multiple Majorana cones, offering a possible microscopic scenario for the anomalous low-temperature longitudinal thermal transport reported in magnetically ordered Kitaev materials such as Na$_2$Co$_2$TeO$_6$. Our results reveal a microscopic route to fractionalized magnetism beyond the conventional dichotomy between magnetic order and spin-liquid behavior.
\end{abstract}

\maketitle

\textbf{\emph{Introduction}}---Frustrated quantum magnets provide a fertile platform for emergent quantum phenomena beyond the conventional Landau paradigm. 
Conventional magnets are characterized by symmetry breaking and local order parameters, whereas quantum spin liquids (QSLs) evade magnetic order 
and exhibit long-range entanglement and fractionalized excitations~\cite{Lee2006-mw,Balents2010-td,Savary2017-bs,Zhou2017-pe,Knolle2019-hn,Broholm2020-pq}. 
An intriguing possibility is that magnetic order can arise without fully confining the fractionalized quasiparticles inherited from a proximate QSL. Such a fractionalized magnet would host long-range magnetic order together with remnant topological structure and deconfined fractionalized excitations.

A natural route is the condensation of local topologically trivial bosonic composites, which can induce magnetic order without the confinement associated with topologically nontrivial anyon condensation~\cite{Bais2009,Neupert2016,Burnell2018,Hwang2024}. 
The emergent gauge structure can then remain intact across the transition.
Related magnetically ordered states with deconfined fractionalized excitations have been discussed in parton mean-field theories~\cite{ZXLiu2010,Savary2012,edrakyan2015,Chern2019,samajdar2019enhanced}, and supported more recently by density-matrix renormalization group calculations~\cite{Huang2022}. 
However, establishing such a phase in a microscopic spin model remains challenging: 
the same interactions that drive magnetic order may also destabilize the underlying gauge structure, making a deconfined magnet difficult to distinguish from a conventional confined ordered state. 
Microscopic settings that both diagnose remnant topological structure and identify the bosonic ordering channel therefore remain scarce.

A paradigmatic QSL is realized in the spin-$1/2$ Kitaev honeycomb model, an exactly solvable $\mathbb{Z}_2$ spin liquid, namely Kitaev spin liquid (KSL), with emergent Majorana fermions coupled to static $\mathbb{Z}_2$ gauge fluxes (visons)~\cite{Kitaev2006-xm}.
Its bond-dependent interactions, naturally arising in spin-orbit-coupled Mott insulators~\cite{Jackeli2009-yn}, have motivated extensive studies of candidate Kitaev materials~\cite{Hermanns2018-vh,Takagi2019-ok,Trebst2022-go,Rousochatzakis2024-td,Matsuda2025-pd}, including honeycomb iridates, $\alpha$-RuCl$_3$, and cobaltates such as Na$_2$Co$_2$TeO$_6$. 
Although most candidate materials develop low-temperature magnetic order due to additional non-Kitaev interactions, 
a variety of experiments have reported signatures suggestive of proximate QSL physics, including broad continua in inelastic neutron scattering~\cite{Banerjee2016,Banerjee2017}
and unusual thermal-transport properties~\cite{Matsuda2018,Matsuda2021,Ong2021,Ong2022}.
Notably, recent longitudinal thermal-transport measurements on Na$_2$Co$_2$TeO$_6$ further suggest that unconventional gapless excitations may
persist inside magnetically ordered regimes~\cite{GangChen2023}. 
This motivates the possibility of a fractionalized antiferromagnet~\cite{Senthil2000,Senthil2003}, where magnetic order coexists with deconfined fractionalized quasiparticles. 
The central question is whether such a coexistence regime, with magnetic order, remnant topological structure, and gapless fractionalized excitations, can be stabilized in a microscopic spin model relevant to Kitaev materials.

Motivated by studies of cobaltate Kitaev materials, where third-neighbor exchange has been argued to be sizable~\cite{Chernyshev2020,Kim2020-px,Choi2012-rp,Singh2012-fm,Kim_2022,Lin2021-tm,Yao2022-cj},
we investigate the Kitaev-$J_3$ model on the honeycomb lattice, in which nearest-neighbor Kitaev interactions compete with an antiferromagnetic (AFM) 
third-nearest-neighbor Heisenberg coupling.
Combining variational Monte Carlo calculations with a vison-quasiparticle analysis near the Kitaev limit,
we provide evidence for deconfined magnetic phases in which $\mathbb{Z}_2$ fractionalization survives inside magnetically ordered states.
Specifically, the deconfined zigzag and AFM states retain multiple linearly independent topological sectors on a torus,
consistent with vison-flux insertion, whereas conventional ordered phases lose this structure. 
We further show that magnetic order can emerge through the condensation of bosonic vison-pair modes,
while single visons remain gapped.
This separation of channels provides a microscopic mechanism by which magnetic order develops without immediate confinement.
Together with reconstructed gapless spinon spectra, these results identify regimes where conventional symmetry breaking coexists with remnant $\mathbb{Z}_2$ topological structure and gapless fractionalized excitations.

\textbf{\emph{Model Hamiltonian}}---The Hamiltonian of the Kitaev-$J_3$ model reads
\begin{align}\label{eq:H_K-J3}
    H = \sum_{\langle i,j \rangle_\gamma} K \, S_i^\gamma S_j^\gamma
    + \sum_{\langle \! \langle \! \langle i,j \rangle \! \rangle \! \rangle} J_3 \, \bm{S}_i \cdot \bm{S}_j,
\end{align}
where $\langle i,j \rangle_{\gamma}$ denotes a nearest-neighbor (NN) $\gamma$-bond ($\gamma = x,y,z$)
while $\langle \! \langle \! \langle i,j \rangle \! \rangle \! \rangle$ denotes a third-NN bond.
We parameterize the couplings by $K=\sin\theta$ and $J_3=\cos\theta$, and focus on the regime $\theta\in[-\pi/2,\pi/2]$.

Besides the spin-orbit-entangled symmetry $D_{3d}\times \mathbb{Z}_2^{\mathcal T}$, with $\mathcal T$ denoting time reversal, the Hamiltonian also possesses pure spin $\pi$-rotation symmetries forming $D_2^{s} = \{ E, C_{2x}, C_{2y}, C_{2z} \}$.
The full symmetry is therefore a spin-space-group symmetry~\cite{Corticelli2022,Fang2024,Song2024,LiuQ2024}, which relates the degenerate magnetic configurations.
The model also admits an exact four-sublattice duality transformation $\scT_4$~\cite{Rousochatzakis2024-td,CChen2026}, which maps $(K,J_3)$ to $(-K,J_3)$, or equivalently $\theta$ to $-\theta$ under this parameterization.

% ------------------------------------------------------------------------------
% FIGURE
% ------------------------------------------------------------------------------
\begin{figure}[t]
\centering
\includegraphics[width = \linewidth]{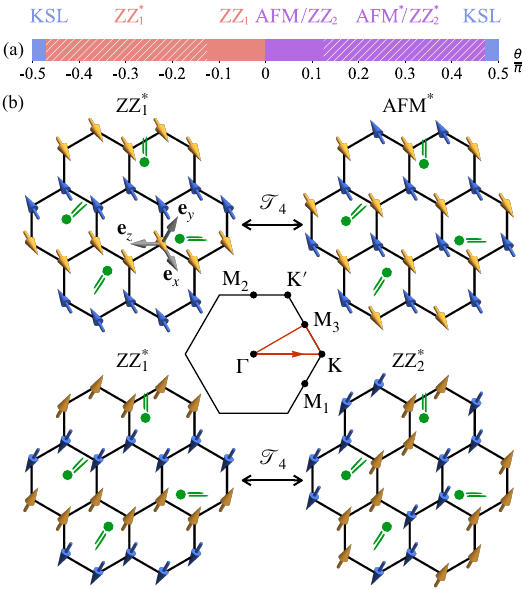}
\caption{(a) The VMC phase diagram of the Kitaev-$J_3$ model, showing the KSL, deconfined zigzag and antiferromagnetic phases ($\mathrm{ZZ}_{1,2}^*$ and $\mathrm{AFM}^*$), and their conventional counterparts ($\mathrm{ZZ}_{1,2}$ and $\mathrm{AFM}$). 
(b) Representative spin configurations in the deconfined magnetic phases.
The two sets of configurations are related by the four-sublattice duality transformation $\mathcal T_4$. 
The green dots schematically denote deconfined itinerant spinon excitations in the fractionalized ordered regimes.
Inset: Brillouin zone with high-symmetry momenta.
}\label{fig:PhaseDiagram}
\end{figure}
% ------------------------------------------------------------------------------

% ------------------------------------------------------------------------------
% FIGURE
% ------------------------------------------------------------------------------
\begin{figure*}[t]
\centering
\includegraphics[width = \linewidth]{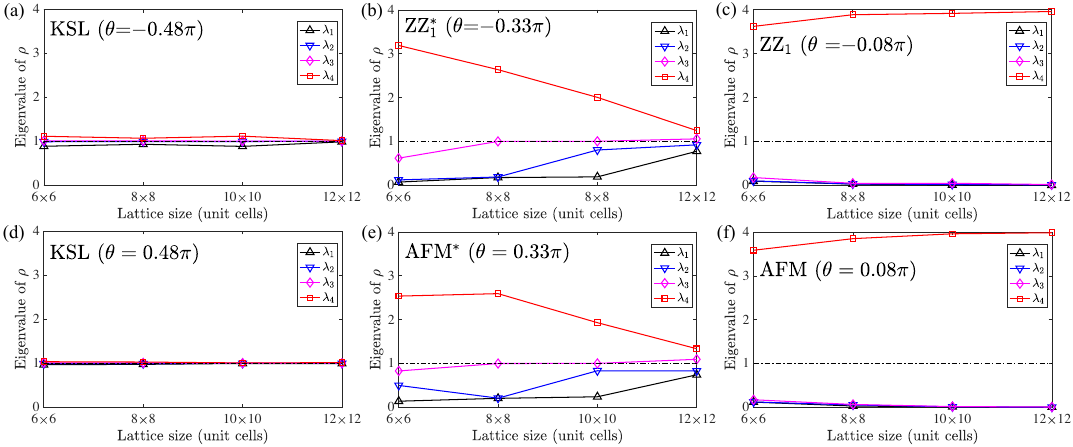}
\caption{Eigenvalues of the normalized overlap matrix $\rho$ on tori of different lattice sizes. 
(a,d) The KSL exhibits four finite eigenvalues, reflecting the fourfold degeneracy of a $\mathbb{Z}_2$ spin liquid.  
(b,e) The deconfined phases ZZ$_1^*$ and AFM$^*$ retain four finite eigenvalues, albeit with enhanced finite-size splitting, indicating remnant $\mathbb{Z}_2$ topological structure coexisting with magnetic order.
(c,f) The conventional ZZ and AFM phases show a single dominant eigenvalue, signaling $\mathbb{Z}_2$ confinement.
}\label{fig:GSD}
\end{figure*}
% ------------------------------------------------------------------------------

\textbf{\emph{VMC Results}}---We first employ variational Monte Carlo (VMC) with Gutzwiller-projected fermionic parton wave functions, in which the spin operators are represented by Abrikosov fermions subject to a single-occupancy constraint enforced by Gutzwiller projector $P_{\rm G}$, as described in the Supplemental Material (SM)~\footnote{See Supplemental Material for details on trial wave functions, order-by-disorder analysis, spinon mean-field dispersions, finite-size criterion for the GSD diagnostic, stability analysis, remarks on entanglement-based diagnostics, $J_3$-induced single-vison and vison-pair hopping processes, and magnetic order from BVPs condensation.}.                          
Rewriting Eq.~(\ref{eq:H_K-J3}) in fermionic operators and decoupling the interactions yields a quadratic mean-field Hamiltonian $H_{\rm mf}(\eta)$, supplemented by site-dependent background fields that capture symmetry-breaking magnetic orders, where $\eta$ collects the variational parameters.
The trial spin wave function $|\Psi(\eta)\rangle=P_{\rm G}|\Psi_{\rm mf}(\eta)\rangle$ is optimized by minimizing its variational energy through Monte Carlo sampling, on tori of size up to $12\times12$ unit cells, corresponding to 288 sites.
All conventional and deconfined states are treated within the same variational framework on equal footing.

The VMC phase diagram of the Kitaev–$J_3$ model is shown in Fig.~\ref{fig:PhaseDiagram}(a). 
Starting from the KSL at $J_3/|K| \approx 0$, increasing the relative strength of $J_3$ drives the system into magnetically ordered phases. 
For the FM Kitaev interaction ($K<0$), the leading instability is toward zigzag order, denoted by ZZ$_1$, with the ordering wave vector 
$\bQ = \M_{1,2,3}$ parallel to a lattice bond (see Fig.~\ref{fig:PhaseDiagram}(b)). Classically, when $\bQ$ is parallel to the $\alpha$ bond, the ordered moments tend to lie in the $\beta$-$\gamma$ plane, 
where $\{\alpha,\beta,\gamma\}$ is a permutation of $\{x,y,z\}$.
The VMC optimization lifts this residual in-plane degeneracy and selects moments aligned along the $\beta$ or $\gamma$ axis,
consistent with an order-by-disorder mechanism~\cite{Rau2018,Rao2026} (see SM~\cite{Note1} for details).
For example, when $\bQ=\M_2$ (parallel to the $z$ bond), the VMC optimization selects symmetry-related zigzag states with moments along the $x$ or $y$ axis [Fig.~\ref{fig:PhaseDiagram}(b)].
On the AFM Kitaev side ($K>0$), the $J_3$ interaction drives the system into either N\'eel AFM order or a distinct zigzag order, denoted by ZZ$_2$. 
Moreover, as illustrated in Fig.~\ref{fig:PhaseDiagram}(b), the AFM and ZZ$_2$ orders are related to ZZ$_1$ by the $\scT_4$, 
confirming that the variational states respect this microscopic duality.

Notably, the ordered states adjacent to the KSL retain signatures inherited from the parent KSL, including topological ground-state degeneracy (GSD) and fractionalized spinon excitations (see below).
We therefore denote these deconfined ordered phases as ZZ$_{1,2}^*$ and AFM$^*$, respectively, in contrast to the conventional magnetically ordered phases ZZ$_{1,2}$ and AFM at larger $J_3$.

To diagnose the remnant $\mathbb{Z}_2$ topological structure, we examine the GSD on a torus~\cite{Wen1989,Wen1990,Kitaev2003}.
Distinct sectors are generated by imposing periodic or antiperiodic boundary conditions for the fermionic spinons along the two noncontractible cycles, labeled by $\mu\in\{++,+-,-+,--\}$.
These sectors represent vison-flux insertions rather than symmetry-related magnetic degeneracies.
For each optimized state, we define the normalized overlap matrix
\begin{equation} \rho_{\mu\nu}= \frac{\langle P_{\rm G}\Psi_\mu|P_{\rm G}\Psi_\nu\rangle} {\sqrt{\langle P_{\rm G}\Psi_\mu|P_{\rm G}\Psi_\mu\rangle \langle P_{\rm G}\Psi_\nu|P_{\rm G}\Psi_\nu\rangle}}, 
\end{equation}
where $|\Psi_\mu \rangle$ denotes the mean-field ground state in sector $\mu$.
With this normalization, $\rho_{\mu\mu}=1$ and $\mathrm{Tr}\rho=4$. 
A deconfined state retains four finite eigenvalues in the thermodynamic limit, whereas a confined magnetic phase exhibits a rank-one structure, with one eigenvalue approaching four and the remaining three vanishing.
We therefore use the ratio $\lambda_1/\lambda_4$, with eigenvalues ordered as $\lambda_1\leq\cdots\leq\lambda_4$, as a finite-size diagnostic of the survival of topological sectors (see SM~\cite{Note1}).

As shown in Fig.~\ref{fig:GSD}, the KSL exhibits the expected fourfold topological degeneracy, while deep in the conventional $\mathrm{ZZ}$ and $\mathrm{AFM}$ phases only one eigenvalue remains finite, indicating confinement of the underlying $\mathbb{Z}_2$ gauge structure.
By contrast, the intermediate deconfined magnetic phases, $\mathrm{ZZ}_{1,2}^{*}$ and $\mathrm{AFM}^{*}$, retain four finite eigenvalues even in the presence of long-range magnetic order, albeit with stronger finite-size splitting than in the KSL. 
The rank-one collapse in conventional ordered phases rules out an interpretation in terms of ordinary symmetry-related magnetic degeneracies and supports the survival of topological sectors in the deconfined magnetic phases.

To further characterize the deconfined magnetic phases, we examine the spinon mean-field spectra of the optimized ansatz (see SM~\cite{Note1}). 
The KSL contains two symmetry-related Majorana cones at low energies. 
Upon magnetic ordering, the coupling between the order parameter and the fractionalized spinons strongly reconstructs the dispersion. 
At the mean-field level, this reconstruction yields multiple Majorana cones: up to fourteen in the $\mathrm{AFM}^{*}$ phase without Brillouin-zone folding, and twelve in the reduced Brillouin zone of the four-sublattice $\mathrm{ZZ}_{1}^{*}$ phase.
Although inversion $\mathcal{P}$ and time reversal $\mathcal{T}$ are individually broken, the combined symmetry $\mathcal{PT}$ remains intact in the optimized ansatz and protects the gapless nodal structure~\cite{Wen2012,PhysRevB.102.094416,Liu2025}. 
The persistence of $\mathcal{PT}$-protected gapless nodes provides a qualitative fingerprint of deconfined magnetism, in contrast to conventional confined magnets where low-energy spinons are absent. 
The remnant $\mathbb Z_2$ topological structure and gapless spinons suggest entanglement-based probes of the gauge structure, with finite-size effects from gapless modes carefully controlled (see SM~\cite{Note1}).
Note that the mean-field dispersions serve as diagnostics rather than directly observable spectra, since Gutzwiller projection and gauge fluctuations can significantly renormalize spectral properties without altering the underlying topological structure.

Moreover, we test the stability of the deconfined magnetic phases against a weak magnetic field (see SM~\cite{Note1}). 
We find that the overlap matrix of the reoptimized variational states continues to exhibit multiple finite eigenvalues throughout the deconfined regimes, 
although their finite-size splittings are modified by the field.
This indicates that the deconfined magnetic phases occupy finite parameter regimes rather than fine-tuned transition lines.

\textbf{\emph{Vison Perspective}}---We further clarify the microscopic origin of the deconfined magnetic phases through 
a vison-quasiparticle analysis of the parent KSL.
Although visons are static in the Kitaev limit, non-Kitaev interactions can endow them with dynamics and drive vison instabilities.
Different instabilities provide distinct routes out of the KSL~\cite{Zhang_2021_vs,Zhang_2022_to,Chen_2023_no,Chen_2025_ap,Chen_2025_apa}: 
single-vison proliferation confines the itinerant spinons and destroys the $\mathbb{Z}_2$ topological structure, whereas condensation of a bosonic vison pair (BVP), 
formed by two neighboring visons as a topologically trivial quasiparticle,
% a topologically trivial two-vison bound state,
can induce symmetry-breaking magnetic order while preserving the underlying $\mathbb{Z}_2$ gauge structure, with the ordering pattern encoded in the soft BVP wave function~\cite{Zhang_2021_vs,Chen_2025_ap,Chen_2025_apa}.
Here, we analyze the $J_3$-induced dynamics of both single visons and BVPs.

% ------------------------------------------------------------------------------
% FIGURE
% ------------------------------------------------------------------------------
\begin{figure}[t]
\centering
\includegraphics[width = 0.4\textwidth]{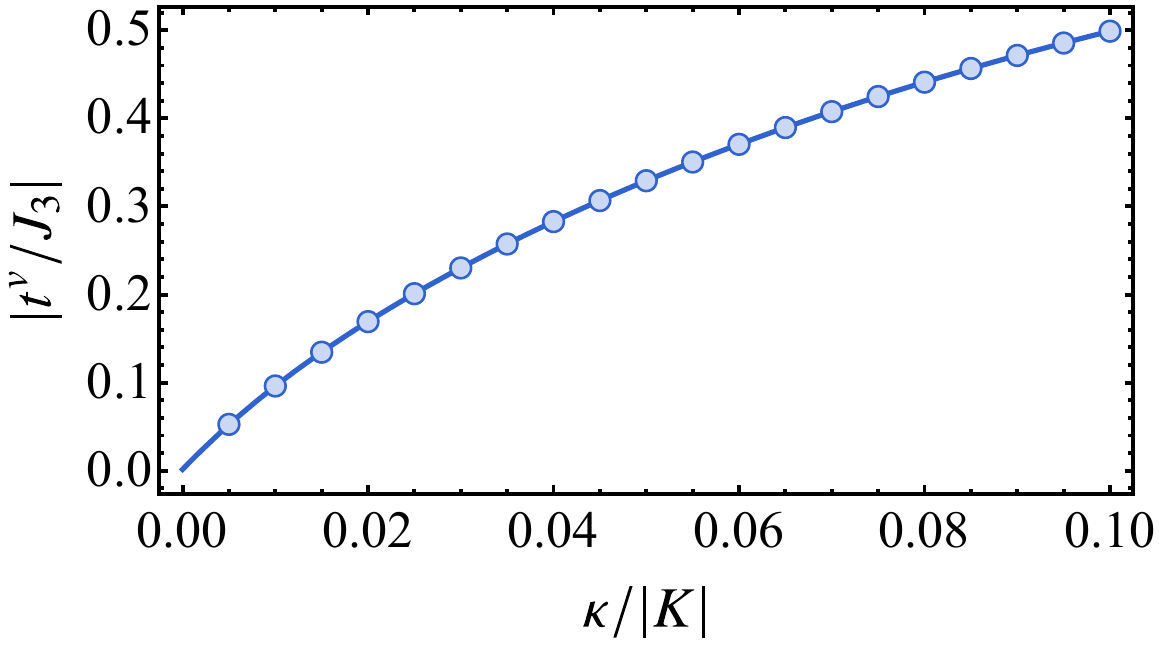}
\caption{Single-vison hopping induced by $J_3$ for FM and AFM Kitaev couplings.
For vison separations exceeding the matter-fermion localization length $\xi_c \propto 1/\kappa$, 
the effective single-vison hopping vanishes as $\kappa\rightarrow0$, 
suggesting the absence of single-vison softening at small $J_3$.
}\label{fig:sv-hopping}
\end{figure}
% ------------------------------------------------------------------------------ 

To construct effective tight-binding descriptions for visons and BVPs, we compute their leading-order hopping amplitudes induced by $J_3$ (see SM~\cite{Note1}).
For single visons, we place two well-separated visons on a torus and evaluate the hopping matrix element $t^{v}_{f,i}\equiv\langle v_f|H_{J_3}|v_i\rangle$, where $|v_i\rangle$ and $|v_f\rangle$ denote the initial and final single-vison states.
To make single visons well-defined local quasiparticles, we introduce a three-spin term ($\kappa$) which generates a Haldane mass for matter Majorana 
fermions~\cite{Kitaev2006-xm} and localizes the fermionic dressing of a vison over the length scale $\xi_c=v_c/\Delta_c$.
Here $v_c=\sqrt{3}|K|$ is the matter-fermion velocity and $\Delta_c=6\sqrt{3}\kappa$ is the induced gap.
Taking the limit $\kappa \rightarrow 0$ while keeping the vison separation larger than $\xi_c$, we find that the leading-order $J_3$-induced 
single-vison hopping vanishes for both FM and AFM Kitaev interactions (see Fig.~\ref{fig:sv-hopping}). 
This suppresses leading-order single-vison softening, consistent with elementary visons remaining gapped over a finite range of $J_3$, while the instability out of the KSL is instead driven by the softening and condensation of BVP modes.

\begin{figure}[t]
\centering
\includegraphics[width = 0.49\textwidth]{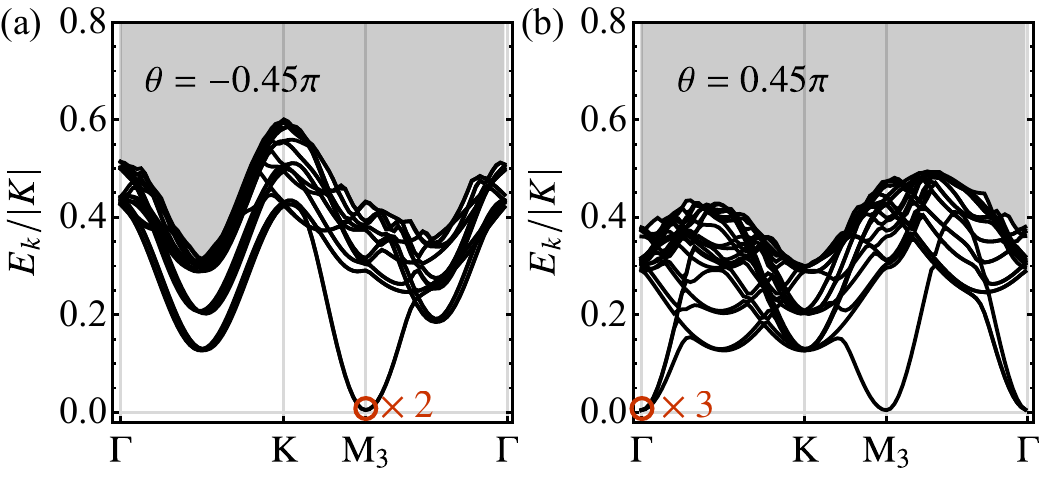}
\caption{BVP dispersions near the magnetic instability. 
(a) On the FM Kitaev side, gap closing occurs at three $\M$ points, each twofold degenerate, leading to the ZZ$_1$ instability.
(b) On the AFM Kitaev side, gap closing occurs at both the $\Gamma$ (threefold degenerate) and $\M$ points, corresponding to N\'eel AFM and ZZ$_2$ instabilities, respectively. 
The dispersions are plotted along the path highlighted in Fig.~\ref{fig:PhaseDiagram}; only the lowest 20 bands are shown.
}\label{fig:vp-bands}
\end{figure}

In contrast to the absence of single-vison hopping, $J_3$ generates finite hopping amplitudes for BVPs, producing dispersive bands that soften near the magnetic instabilities.
By analyzing the corresponding soft-mode wave functions, we determine the magnetic ordering pattern induced by BVP condensation (see SM~\cite{Note1}).
For the FM Kitaev interaction, near $\theta\simeq -0.45\pi$, two degenerate bosonic modes soften at each $\M$ point (see Fig.~\ref{fig:vp-bands}(a)).
Their condensation generates ZZ$_1$ order, with moments along two distinct spin axes, consistent with the ZZ$_1$ pattern from VMC (see Fig.~\ref{fig:PhaseDiagram}(b)).
For the AFM Kitaev interaction, the dual instability occurs near $\theta\simeq 0.45\pi$, where the BVP gap closes at both the $\Gamma$ and $\M$ points (see \cref{fig:vp-bands}(b)). 
The three degenerate soft modes at $\Gamma$ induce N\'eel AFM order with the moments aligned along the three spin axes, whereas the soft modes at the $\M$ points produce ordering patterns consistent with ZZ$_2$.
Thus, the magnetic orders induced by BVP condensation on the FM and AFM Kitaev sides agree with those obtained from VMC calculations and are connected by the $\scT_4$ duality transformation.
The BVP softening scale $|\theta|\simeq0.45\pi$ is close to the VMC KSL boundary at $|\theta|\simeq0.47\pi$, supporting consistency in both symmetry and energy scale.

Since each BVP carries a trivial $\mathbb Z_2$ flux, its condensation does not directly lead to proliferation of elementary visons.
Together with the suppression of single-vison hopping, this supports a deconfined magnetic regime in which magnetic order develops through the BVP channel while elementary vison softening is avoided.
The BVP softening suggests a continuous instability; the associated critical theory, including its coupling to gapless spinons, is left for future work.

% ------------------------------------------------------------------------------
% FIGURE
% ------------------------------------------------------------------------------
% ------------------------------------------------------------------------------

\textbf{\emph{Discussion and Conclusion---}}Our results provide evidence that the Kitaev-$J_3$ model realizes a microscopic setting for deconfined magnetic phases, 
where long-range magnetic order coexists with remnant $\mathbb{Z}_2$ topological structure and fractionalized quasiparticles.
The central evidence is the persistence of nontrivial ground-state degeneracy on a torus: the deconfined ZZ$_{1,2}^*$ and AFM$^*$ phases 
retain multiple linearly independent topological sectors, whereas conventional ordered phases collapse to a single sector. 
This structure further survives a weak symmetry-lowering magnetic field, indicating that the coexistence of magnetic order and topological structure is not fine-tuned. 
Microscopically, the stability is supported by a separation of vison channels: magnetic order develops through a topologically trivial BVP channel, whereas leading-order single-vison softening is suppressed. 
Thus magnetic order can emerge without immediate proliferation of elementary visons,  allowing fractionalized excitations and the remnant $\mathbb{Z}_2$ 
gauge structure to survive within an ordered magnetic background. 
The magnetic ordering patterns and the KSL boundary are broadly consistent with previous numerical and mean-field-plus-RPA studies~\cite{Kim2020-px,CChen2026},
while the present work further shows that the intermediate ordered regimes retain remnant $\mathbb Z_2$ topological structure.

The reconstructed spinon spectra characterize the low-energy excitations of these phases.
At the mean-field level, the spinon bands remain gapless with multiple Majorana cones; these gapless quasiparticles may remain stable against weak gauge fluctuations and could contribute to low-temperature thermal transport, offering a possible microscopic scenario for the anomalous longitudinal thermal conductivity observed in Na$_2$Co$_2$TeO$_6$~\cite{GangChen2023}. 
Moreover, the magnetic background hybridizes different fractionalized quasiparticles of Kitaev's parton description, suggesting that the spin dynamical structure factor may exhibit a gapless spinon continuum coexisting with magnon-like modes associated 
with magnetic order~\cite{Wang2019}.
Such a low-energy continuum would provide a characteristic neutron-scattering signature of the deconfined magnetic phases, although a quantitative dynamical-VMC calculation is left for future work~\cite{Li2010,LiuZX2021,Becca2019}.

This perspective may extend to more realistic models with additional $\Gamma$ and $\Gamma'$ interactions, where candidate fractionalized ordered regimes were reported near the AFM KSL for negative $\Gamma$~\cite{Chen_2025_ap}. 
More generally, if the parent QSL is not adiabatically connected to the Kitaev limit, the vison-pair description must be replaced by the appropriate topologically trivial composite mode determined by the quasiparticle content of that QSL.

In summary, the VMC and vison-quasiparticle analyses together provide evidence for a microscopic route to fractionalized magnetism: the intermediate ordered regimes retain nontrivial topological sectors on a torus while magnetic order emerges through a topologically trivial BVP channel.
The nature of the eventual confinement transition into conventional ordered phases at larger $J_3$, whether sharp or crossover-like, remains an open question. 
More broadly, our results suggest that magnetic order in Kitaev materials need not preclude deconfined low-energy quasiparticles.

{\it Acknowledgments}---We thank Zheng-Xin Liu for insightful discussions and valuable comments.
J.W. acknowledges support from the National Natural Science Foundation of China under Grant No.~12404170 and the start-up grant at HZNU.
C.C. acknowledges support from the National Natural Science Foundation of China (Grants No.~12404175 and No.~12247101), the Fundamental Research Funds for the Central Universities (Grant No.~lzujbky-2025-jdzx07), the Natural Science Foundation of Gansu Province (No.~22JR5RA389 and No.~25JRRA799).

\bibliography{reference.bib}

\clearpage
\widetext

\renewcommand{\theequation}{S\arabic{equation}}
\setcounter{equation}{0}
\renewcommand{\thefigure}{S\arabic{figure}}
\setcounter{figure}{0}
\renewcommand{\thetable}{S\arabic{table}}
\setcounter{table}{0}

%==appendix==
\section{Supplemental Material}

\subsection{A. Construction of trial wave functions}\label{VMC}
The variational Monte Carlo (VMC) approach used here is based on Gutzwiller-projected trial wave functions, which provide a powerful framework for studying strongly correlated quantum systems, particularly quantum spin liquids.
We use the Abrikosov-fermion representation, where spin operators are expressed in terms of fermionic spinons as 
\begin{equation}
S_i^m =\frac{1}{2} C_i^\dagger \sigma^m C_i, 
\end{equation}
where $C_i^\dagger = (c_{i\uparrow}^\dagger,c_{i\downarrow}^\dagger)$, $m \equiv x,y,z$, and $\sigma^m$ are Pauli matrices.
To restrict the Hilbert space to physical spin states, the single-occupancy constraint $\hat{N_i} = c_{i\uparrow}^\dagger c_{i\uparrow} + c_{i\downarrow}^\dagger c_{i\downarrow} = 1$ must be enforced on every lattice site.
It is convenient to introduce the matrix operator $\psi_i=( C_i, \bar C_i)$ with $\bar C_i=(c_{i\dn}^\dag, -c_{i\up}^\dag)^T$ such that the spin operators can also be written as $S^m = {\rm Tr}(\psi_i ^\dag {\sigma^m \over4}\psi_i)$.
This representation makes explicit both the global SU(2) spin-rotation symmetry and an independent local SU(2) gauge redundancy.

Meanwhile, the two spinon species may further be expressed in terms
of four Majorana fermions,
\begin{equation}\label{Rep}
c_{i\up} = {\textstyle \frac12} (b_i^z + ic_i),\ \ \
c_{i\dn} = {\textstyle \frac12} (b_i^x + ib_i^y),
\end{equation}
which satisfy the anti-commutation relations $\{b_i^\alpha, b_j^\beta\} = 2
\delta^{\alpha \beta}\delta_{ij}$ ($\alpha,\beta = 0,x,y,z$; $b^0 \equiv c$). In this
basis, the spin operator takes the form
\begin{equation}
S_i^m = \frac{i}{2}b_i^m c_i,
\end{equation}
and the particle-number constraint is $b_i^x b_i^y b_i^z c_i = 1$.
Then the matrix operator can equivalently be written as $\psi_i =  {\textstyle \frac12} (ic_i + b_i^x \tau_x + b_i^y \tau_y + b_i^z \tau_z)$. 
Furthermore, the particle-number constraint is ${\rm Tr}(\psi_i \tau^m \psi_i^\dagger)=0, m=x,y,z$,
where we use the Pauli matrix $\tau^m$ to distinguish from the spin operation
$\sigma^m$.

In the VMC framework, the spin interactions are rewritten in terms of interacting fermionic operators and subsequently decoupled into a quadratic mean-field Hamiltonian $H_{\rm mf}(\eta)$, where $\eta$ denotes a set of variational parameters specified below.
The trial wave function is then obtained by applying the Gutzwiller projection to the mean-field ground state $|\Psi (\eta)\rangle = P_{\rm G} |\Psi_{\rm mf}(\eta) \rangle$, with $P_{\rm G}$ enforcing the single-occupancy constraint. 
The variational energy $E(\eta) = \langle \Psi(\eta) |H| \Psi(\eta) \rangle / \langle \Psi(\eta)| \Psi(\eta) \rangle$ is evaluated by Monte Carlo sampling, and the optimal variational parameters $\eta$ are determined by minimizing energy.
In the following we describe the construction of the trial mean-field Hamiltonian in detail.

To decouple Kitaev interactions in the Kitaev-$J_3$ model, we employ a “Kitaev-type” mean-field ansatz commonly adopted in previous studies,
\begin{equation}\label{Kitaevmf}
H_{\rm mf}^{\rm K}  =  \sum_{\langle i,j \rangle_{\gamma}}  \rho_a (ic_ic_j) +
\rho_c (i b_i^\gamma b_j^\gamma).
\end{equation}
In the Abrikosov-fermion notation, this ansatz can be written equivalently as
\begin{align}
H_{\rm mf}^{\rm K}
=& \sum_{\langle i,j\rangle_{\gamma}}
\Bigg[
i\rho_a\,{\rm Tr}\!\left(
\psi_i^\dagger\psi_j
+\tau^x\psi_i^\dagger\sigma^x\psi_j
+\tau^y\psi_i^\dagger\sigma^y\psi_j
+\tau^z\psi_i^\dagger\sigma^z\psi_j
\right) \nonumber \\
&\qquad\qquad
+ i\rho_c\,{\rm Tr}\!\left(
\psi_i^\dagger\psi_j
+\tau^\gamma\psi_i^\dagger\sigma^\gamma\psi_j
-\tau^\alpha\psi_i^\dagger\sigma^\alpha\psi_j
-\tau^\beta\psi_i^\dagger\sigma^\beta\psi_j
\right)
+{\rm H.c.}
\Bigg].
\end{align}
where $\{\alpha,\beta,\gamma\}$ is a permutation of $\{x,y,z\}$.
$\rho_a$ and $\rho_c$ are both real numbers.
Meanwhile, the mean-field Hamiltonian $H_{\rm mf}^{\rm K}$ preserves all symmetries of the original system through the projective symmetry group (PSG)~\cite{Wen2002,You2012}.

Generally, the third-NN AFM Heisenberg interaction $J_3$ is decoupled into the quadratic form
\beq
H_{\rm mf}^{\rm J_3} = \sum_{\langle\langle\langle i,j \rangle\rangle\rangle} {\rm Tr} \!\left( U_{ij}^\dag \psi_i^\dag \psi_j + {\rm H.c.} \right) + \sum_i \pmb \lambda_i \! \cdot \! \pmb \Lambda_i,
\eeq
where $\pmb \lambda$ are the Lagrange multipliers enforcing the local constraints $\Lambda_i^m =  {\rm Tr} (\psi_i \frac{\tau^m}{4} \psi_i^{\dag})$, $m \equiv x,y,z$, and $\tau^m$ are Pauli matrices. The matrix field $U_{ij}^\dag = U_{ji} = \langle \psi_j^\dag \psi_i \rangle$ contains the singlet hopping and pairing channels,
\begin{eqnarray}
\langle \psi_{j}^{\dagger} \psi_{i} \rangle = \left( \begin{matrix} \langle C_{j}^{\dagger} C_{i} \rangle & \langle C_{j}^{\dagger}\bar{C}_{i} \rangle \\ \langle \bar{C}_{j}^{\dagger} C_{i} \rangle & \langle \bar{C}_{j}^{\dagger} \bar{C}_{i} \rangle \end{matrix} \right) = \left( \begin{matrix} \chi_{ji} &
\Delta_{ji}^{*} \\ \Delta_{ji} & -\chi_{ji} \end{matrix} \right).
\end{eqnarray}
Here, $\chi_{ji}$ and $\Delta_{ji}$ denote the singlet hopping and pairing amplitudes, respectively.
In principle, the pairing channel $\Delta_{ji}$ can be included in the mean-field decoupling; however, it generally breaks the PSG structure associated with the Kitaev spin liquid and is therefore omitted in the present work~\cite{Wang2019,Wang2020,Wang2023}. Meanwhile, the hopping amplitude $\chi_{ji}$ is taken to be purely imaginary.

To capture symmetry-breaking magnetic order in the Kitaev-$J_3$ model, we further introduce a site-dependent background field $\pmb M_i$ within the single-$\bQ$ approach~\cite{Kee2014},
\beq\label{Order}
H_{\rm mf}^{\rm Order}=- \sum_i \left(\pmb {M}_i \cdot C_i^\dagger \frac{\pmb\sigma}{2} C_i + {\rm H.c.} \right).
\eeq

Together, the full trial mean-field Hamiltonian in the VMC framework is given by,
\beq
H_{\rm mf}^{\rm Total} = H_{\rm mf}^{\rm K} + H_{\rm mf}^{\rm J_3} + H_{\rm mf}^{\rm Order}.
\eeq
The complete set of variational parameters ($\eta$) includes ($\rho_a$, $\rho_c$, $\chi$, $\pmb\lambda$, $\pmb M_i$). 
The corresponding trial wave function is then obtained by applying the Gutzwiller projection to the mean-field ground state $|\Psi (\eta)\rangle = P_{\rm G} |\Psi_{\rm mf}(\eta) \rangle$, where $P_{\rm G}$ denotes a Gutzwiller projection that enforces the single-occupancy constraint.
The optimal values of variational parameters are determined by minimizing the energy of the Gutzwiller-projected wave function.

Although variational accuracy may be further improved by enlarging the parameter space of the trial wave function such as two-body Jastrow factors, previous VMC studies of spin-orbit-coupled spin models found only negligible effects on the phase boundaries~\cite{Wang2024}. 
Consistently, we observe only very small energy improvements upon including Jastrow correlations, and therefore we omit them in the present VMC calculations.\\

% ------------------------------------------------------------------------------
% FIGURE
% ------------------------------------------------------------------------------
\begin{figure}[b]
\centering
\includegraphics[width = 7.5cm]{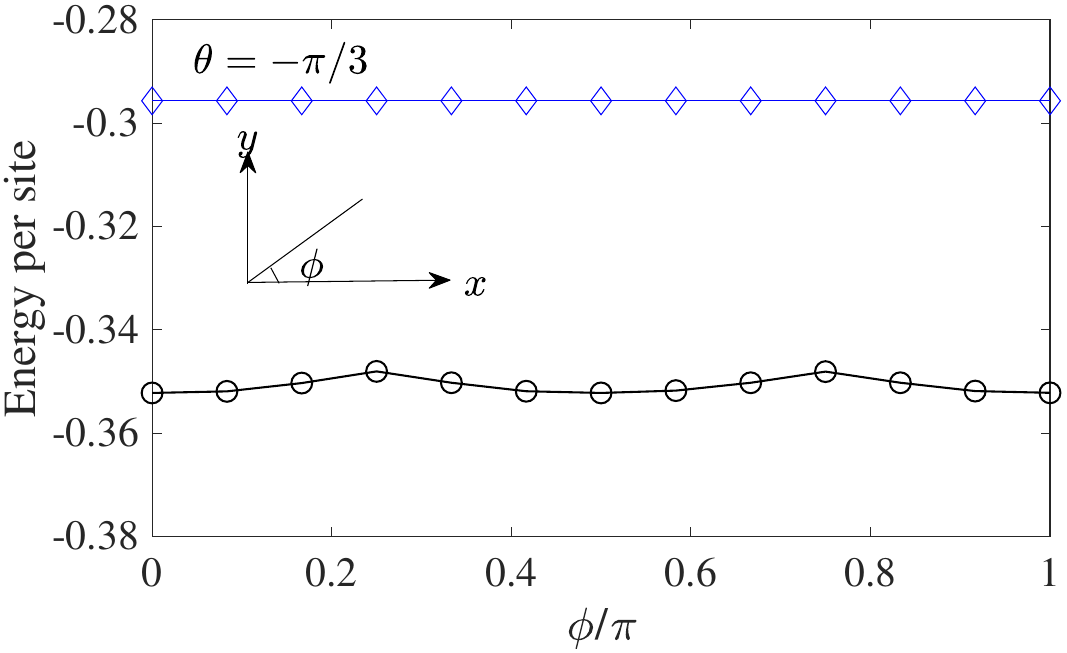}
\caption{Energy per site of the ZZ$_1^*$ state as a function of the spin direction in the $xy$-plane at $\theta=-\pi/3$.
Blue diamonds denote the classical single-$\bQ$ energy, while black circles denote the optimized VMC energy on an $8\times8\times2$ lattice. 
The error bars of the VMC energies are smaller than the symbol size and are therefore not shown.
The VMC minima indicate order-by-disorder selection of discrete spin orientations.}\label{fig:OdO}
\end{figure}
% ------------------------------------------------------------------------------

\subsection{B. Order-by-disorder mechanism}\label{OdO}
In frustrated magnets, competing interactions often give rise to an extensively or accidentally degenerate manifold of classical ground states. 
Although these configurations are degenerate at the classical level, quantum fluctuations can lift the degeneracy and select particular ordering patterns through the order-by-disorder mechanism.
When the continuous degeneracy is accidental rather than symmetry protected, the associated low-energy collective excitations correspond to pseudo-Goldstone modes instead of true Goldstone modes.

In the Kitaev-$J_3$ model, the competition between the bond-dependent Kitaev interaction and the third-nearest-neighbor Heisenberg coupling generates a continuously degenerate manifold of classical magnetic states. 
For example, one representative zigzag configuration consists of spins aligned ferromagnetically along the $x$- and $y$-type bonds, while neighboring spins connected by the z-type bonds are antiferromagnetically aligned. 
At the classical level, equivalently within the classical single-$\bQ$ approximation, the energy is minimized when the ordered moments point along an arbitrary direction within the $xy$-plane, resulting in an accidental continuous degeneracy of the manifold. 
The VMC calculations show that this accidental degeneracy is lifted by quantum fluctuations.
In the ZZ$^*$ or ZZ ordered regime, for example, the variational energy develops distinct minima at several discrete symmetry-related orientations, with the ordered moments preferentially aligned along the $x$ or $y$ directions (and their time-reversed counterparts), as shown in Fig.~\ref{fig:OdO}. For the representative $\bQ=\M_2$ zigzag state, these two selected configurations are related by a spin-space-group operation $\mathcal T C_{2{\rm b}}$, where $C_{2{\rm b}}$ denotes a $\pi$-rotation about the $[-1,1,0]$ direction acting simultaneously in real and spin space, and $\mathcal T$ is time reversal. This operation leaves the $M_2$ ordering channel invariant while interchanging the $x$ and $y$ spin-bond labels. Thus, the VMC optimization lifts the accidental continuous degeneracy of the classical single-$\bQ$ manifold, whereas the remaining discrete degeneracy is symmetry related. The associated pseudo-Goldstone modes therefore acquire a finite gap from the fluctuation-induced anisotropy. This provides a sharp contrast to the gapless fractionalized spinon excitations in the deconfined magnetic phases.

% ------------------------------------------------------------------------------
% FIGURE
% ------------------------------------------------------------------------------
\begin{figure*}[b]
\centering
\includegraphics[width = \linewidth]{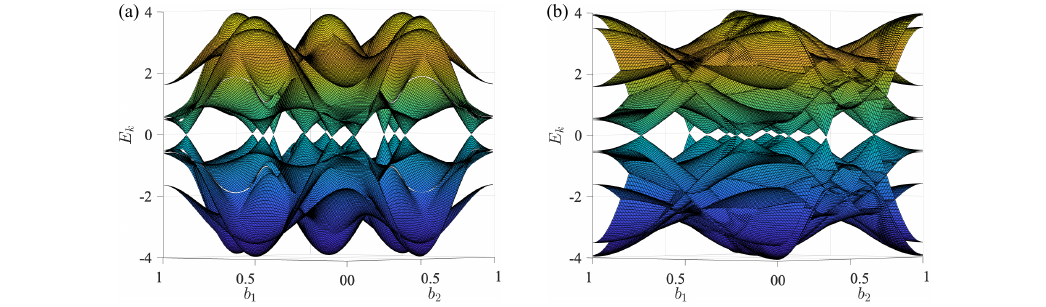}
\caption{Spinon mean-field dispersions in the deconfined magnetic phases. 
(a) Spinon dispersion in the AFM$^*$ at $\theta=\pi/3$, showing fourteen Majorana cones. 
The reciprocal lattice vectors are $\mathbf{b}_1=(-2\pi,\,2\pi/\sqrt{3})$ and $\mathbf{b}_2=(2\pi,\,2\pi/\sqrt{3})$, corresponding to the real-space primitive vectors $\mathbf{a}_1=(-1/2,\,\sqrt{3}/2)$ and $\mathbf{a}_2=(1/2,\,\sqrt{3}/2)$. 
(b) Spinon mean-field dispersion in the ZZ$_1^*$ at $\theta=-\pi/3$, showing twelve Majorana cones. 
Here $\mathbf{b}_1=(2\pi,\,0)$ and $\mathbf{b}_2=(0,\,2\pi/\sqrt{3})$ correspond to $\mathbf{a}_1=(1,\,0)$ and $\mathbf{a}_2=(0,\,\sqrt{3})$.
}\label{fig:dispersion}
\end{figure*}
% ------------------------------------------------------------------------------

\subsection{C. Spinon dispersion of the deconfined magnetic phases}\label{MeanField}
To further characterize the deconfined magnetic phases identified in the main text, we analyze the spinon mean-field spectra of the optimized variational states. 
The mean-field Hamiltonian is fixed by the VMC-optimized variational parameters, and the spectra are evaluated at the mean-field level on larger lattices.
Although the physical excitations after Gutzwiller projection are strongly interacting and subject to gauge fluctuations, the fermionic mean-field dispersion nevertheless provides a useful diagnostic of the remnant fractionalized structure.
Compared with the parent KSL (with two symmetry-related Majorana cones), the deconfined magnetic phases exhibit a strong reconstruction of the spinon spectrum. 
This reconstruction is driven by two ingredients in the optimized ansatz: the $J_3$-induced singlet hopping amplitude $\chi$, which disperses and hybridizes the spinon bands beyond the pure Kitaev limit, and the coupling between the background magnetic order and the fractionalized spinons through the site-dependent field $\pmb M_i$.

In the AFM$^\ast$ phase, the magnetic order preserves the two-sublattice structure of the honeycomb lattice, and therefore does not induce Brillouin-zone folding. 
Nevertheless, the finite $\chi$ and the AFM background field strongly reconstruct the low-energy spectrum, producing as many as fourteen Majorana cones [see Fig.~\ref{fig:dispersion}(a)]. 
Thus the proliferation of cones is not a simple consequence of zone folding, but mainly results from the combined effect of the $J_3$-induced spinon hopping and the hybridization between magnetic order and spinon degrees of freedom.

By contrast, the ZZ$^*$ phase enlarges the magnetic unit cell to four sublattices. 
In this case, Brillouin-zone folding further acts on top of the same microscopic reconstruction mechanism. 
The resulting spinon spectrum contains twelve Majorana cones distributed throughout the reduced Brillouin zone [see Fig.~\ref{fig:dispersion}(b)].

In both deconfined magnetic phases, $\mathrm{ZZ}^{*}$ and $\mathrm{AFM}^{*}$, the finite magnetic background field strongly hybridizes the $c$ and $b^{m}$ Majorana sectors over the full Brillouin zone; see Eqs.~(\ref{Rep}) and (\ref{Order}). 
This hybridization gives the gapless spinons finite overlap with physical spin fluctuations~\cite{Wang2019}. Consequently, spin dynamical structure factor, $S(\mathbf q,\omega)$, is expected to contain a gapless continuum from intra-cone and inter-cone spinon processes, coexisting with gapped magnon-like branches associated with the fluctuation-selected magnetic order. This provides a neutron-scattering signature of the deconfined magnetic phases.

The Gutzwiller projection may renormalize spectral weights and velocities, while the stability of the symmetry-protected nodal structure is controlled by the symmetries retained by the projected state.
The survival of multiple low-energy spinon cones distinguishes these deconfined magnetic phases from conventional confined magnets. 
In a confined magnet, fractionalized spinons are removed from the low-energy spectrum, and magnetic ordering leads instead to conventional magnon excitations. 
Here, by contrast, the reconstructed but still gapless spinon spectrum provides a mean-field fingerprint of deconfined magnetism coexisting with long-range magnetic order.

\subsection{D. Finite-size criterion for the GSD diagnostic} 
For the normalized overlap matrix $\rho$ used in the main text, $\rho_{\mu\mu}=1$ and $\mathrm{Tr} \rho = 4$. Let $0\le\lambda_1\le\lambda_2\le\lambda_3\le\lambda_4$ denote its four eigenvalues with $\sum_i\lambda_i=4$. 
In an ideal deconfined state with four topological sectors, the projected states generated by the four boundary conditions become mutually orthogonal in the thermodynamic limit, so that all four eigenvalues approach one. 
By contrast, in a confined magnetic phase, the four boundary-condition states collapse onto a single physical state, so that $\rho$ becomes rank one, with $\lambda_4\to4$ and $\lambda_{1,2,3}\to0$.

In practice, we monitor the ratio $\lambda_1/\lambda_4$, as a function of the size of the system. 
This ratio approaches one in a fully deconfined state and zero in a confined state, and therefore serves as a convenient order-parameter-like diagnostic. 
A value of $\lambda_1/\lambda_4$ that saturates to a finite value with increasing system size signals the survival of distinct topological sectors, whereas a rapid decrease towards zero signals their collapse. At the accessible system sizes, this criterion cannot sharply distinguish a true confinement transition from a rapid crossover; the phase boundaries in Fig.~\ref{fig:PhaseDiagram}(a) in the main text should therefore be understood as finite-size estimates of the locus where the remnant topological sectors collapse within our numerical resolution.

\subsection{E. Stability of the deconfined magnetic phases}\label{Stability}
In the main text, the deconfined magnetic phases are identified by the coexistence of long-range magnetic order and nontrivial ground-state degeneracy (GSD) on a torus. 
A natural question is whether these phases are stable against weak perturbations, or instead appear only as a fine-tuned critical line between the KSL and conventional long-range magnetic orders. 
To address this issue, we examine the response of the deconfined magnetic phases to a weak external magnetic field.

For simplicity, we consider the Kitaev-$J_3$ model in a magnetic field applied along the $z$-direction,
\begin{equation}\label{eq:field_hamilt}
H^{\rm Zeeman} =  h_z \sum_i  S_i^z.
\end{equation}
Within the fermionic parton representation, this term is incorporated at the mean-field level as
\begin{equation}
H_{\rm mf}^{\rm Zeeman} = h_z \sum_i  C_i^\dagger \frac{\sigma^z}{2} C_i.
\end{equation}
For each value of $h_z$, the variational wave function is reoptimized using the same VMC framework. 
We then repeat the overlap-matrix analysis on a torus. 
As in the zero-field case, the number of nonvanishing eigenvalues of $\rho$ serves as a diagnostic of the GSD and provides evidence for the survival of the deconfined $\mathbb{Z}_2$ gauge structure.

\begin{figure*}[t]
\centering
\includegraphics[width = \linewidth]{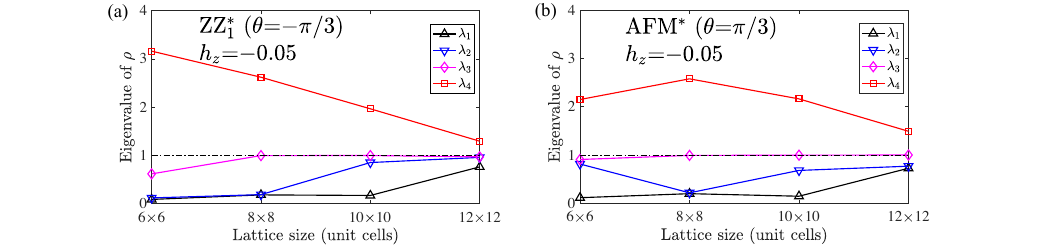}
\caption{Eigenvalues of the normalized overlap matrices $\rho$ on tori of different sizes under a weak $z$-direction magnetic field ($h_z=-0.05$).
In both ZZ$_1^*$ (a) and AFM$^*$ (b), four finite eigenvalues persist with increasing system size, indicating that remnant topological degeneracy coexists with magnetic order and remains stable against weak field perturbations. 
}\label{fig:GSD_field}
\end{figure*}

Our VMC calculations show that the deconfined magnetic phases retain nontrivial topological-sector structure under a weak $z$-direction field such as $h_z=-0.05$. 
Although the field explicitly reduces the symmetry and modifies the optimized magnetic background, the overlap matrix still exhibits multiple finite eigenvalues, as shown in Fig.~\ref{fig:GSD_field}. 
In the small-field regime studied here, the four-sector structure found at $h_z=0$ remains visible, albeit with stronger finite-size splitting than in the KSL. 
This slower convergence may reflect enhanced finite-size hybridization among topological sectors in the magnetically ordered background, possibly mediated by coupling to order-parameter fluctuations.

% ------------------------------------------------------------------------------
% FIGURE
% ------------------------------------------------------------------------------
% ------------------------------------------------------------------------------

This result provides an important stability check for the deconfined magnetic phases. 
If the nontrivial GSD were merely a finite-size remnant of the nearby KSL, or a fine-tuned feature restricted to a phase boundary, it would be expected to disappear rapidly once a symmetry-lowering perturbation such as $h_z$ is introduced. 
Instead, the survival of nontrivial GSD over a finite range of weak magnetic fields indicates that the deconfined magnetic phases occupy finite regions of parameter space. 
This stability can be understood from the underlying $\mathbb Z_2$ gauge structure: as long as the single-vison gap remains finite, weak local perturbations, including additional non-Kitaev interactions such as $\Gamma$ or $\Gamma'$, are not expected to immediately mix or destroy the distinct topological sectors.
They should therefore be understood as stable phases, rather than as isolated critical lines separating the KSL from conventional magnetic orders.
Beyond the weak-field stability addressed here, fields that break the symmetry protecting the Majorana cones may gap the spinons. 
If the induced mass carries a nonzero Majorana Chern number, this could lead to a field-induced chiral fractionalized magnet with coexisting magnetic order, a possibility left for future work.

\subsection{F. Remarks on entanglement-based diagnostics}
The deconfined magnetic phases found here may, in principle, be characterized by entanglement-based diagnostics. 
For gapped topologically ordered phases, the topological entanglement entropy (TEE) for noncontractible bipartitions of a torus depends on the ground-state sector, and the corresponding minimum entropy states (MES) provide a route to extract modular matrices and quasiparticle braiding statistics from the ground-state manifold~\cite{Zhang2012}. 
For the pure KSL, where Majorana fermions are coupled to a static $\mathbb Z_2$ gauge field, the entanglement entropy contains both gauge-field and Majorana-fermion contributions. 
The gauge sector yields the universal topological term $\gamma=\ln2$, whereas the Majorana sector contributes the nonuniversal boundary-law term and controls the low-lying entanglement spectrum~\cite{Yao2010}.
This observation is particularly relevant to the deconfined magnetic phases studied here, where gapless spinons are coupled to a deconfined $\mathbb Z_2$ gauge field. 
Although gapless spinons generate additional finite-size and geometry-dependent contributions to the entanglement entropy, the gapped vison sector is may retain a residual $\mathbb Z_2$ topological contribution. 
Therefore, one may attempt to extract the gauge-sector TEE and MES-like structure by computing the R\'enyi entropy for linear combinations of projected states in different vison-flux sectors.

In practice, however, this program requires a separate large-scale calculation beyond the scope of the present work. 
One needs to evaluate R'enyi entropies for many linear combinations of nearly degenerate projected states and for several noncontractible bipartitions, followed by a systematic finite-size analysis. 
This is particularly demanding here because the gapless spinon cones produce strong finite-size contributions to the entanglement entropy, which must be separated from the residual gauge-sector contribution. 
Moreover, a full modular-matrix characterization is not strictly universal in the presence of gapless modes, since the low-energy theory is not a purely gapped topological quantum field theory. 
We therefore leave a quantitative analysis of the TEE, MES structure, and possible braiding signatures in these gapless deconfined magnetic phases for future work.

\subsection{G. $J_3$-induced single-vison and vison-pair hoppings}
\begin{figure*}[t]
\centering
\includegraphics[width = 0.6\textwidth]{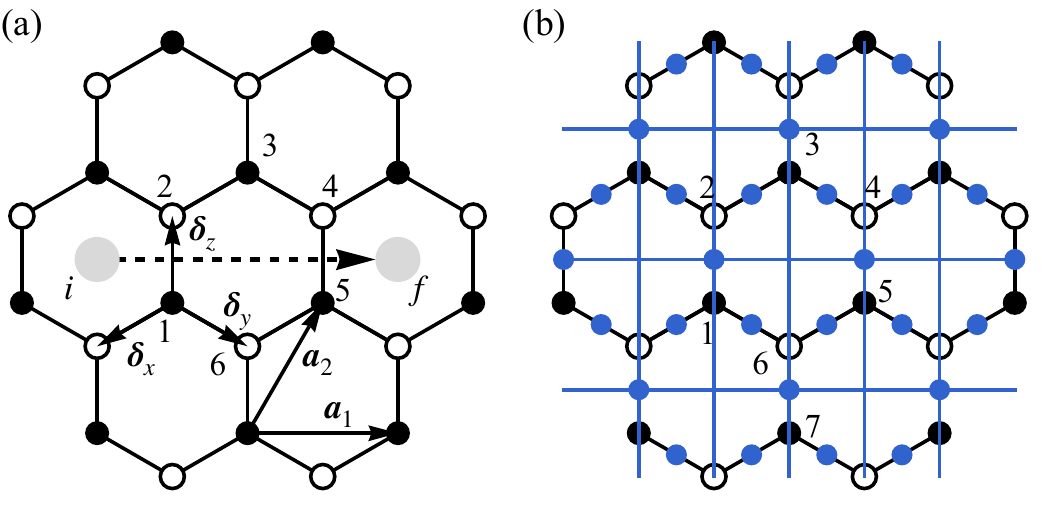}
\caption{
(a) Hopping of a single vison induced by the $J_3$ interaction. For the hopping process illustrated here, the $\sigma_2^z\sigma_5^z$ and $\sigma_1^z\sigma_4^z$ 
terms contribute to the effective hopping amplitude. 
(b) Effective lattice model for BVPs (blue dots).
Here we show the hopping processes of BVPs on $z$ bonds only; the corresponding models for BVPs on $x$ and $y$ bonds are related by $C_3$ rotations.
Black and white dots denote the $A$ and $B$ sublattice sites, respectively.
}\label{fig:vison-move}
\end{figure*}

The $J_3$ interaction can induce single-vison hopping to a third-neighbor plaquette on the triangular lattice formed by the plaquette centers. 
The leading-order $J_3$-induced single-vison hopping amplitude is defined as
$t_{f,i} \equiv \langle v_f | H_{J_3} | v_i \rangle$,
where $|v_i\rangle$ and $|v_f\rangle$ denote the initial and final single-vison states, respectively (see \cref{fig:vison-move}(a)). 
The effective Hamiltonian for a single vison has the form:
\begin{align}
    H_\mathrm{v} = \sum_{i,j} t_{i,j} | v_i \rangle \langle v_j | + \sum_i \Delta_v |v_i \rangle \langle v_i |,
\end{align}
where $\Delta_v \approx 0.15 |K|$ is the excitation energy of a vison in the pure Kitaev model.
For each hopping process, two terms in the $J_3$ interaction contribute to this matrix element. For example, for the hopping process shown in 
\cref{fig:vison-move}(a), both $\sigma_1^z \sigma_4^z$ and $\sigma_2^z \sigma_5^z$ contribute.

Since the calculation is performed on a torus, only configurations with an even number of visons are allowed. We therefore introduce a second vison
far away from the target vison and compute the hopping of the latter in this two-vison background. To make the single vison a well-defined local quasiparticle,
we add a weak three-spin term $\kappa$ to the Hamiltonian. This term gaps the matter Majorana fermions and localizes the matter-fermion dressing of 
a vison over the length scale $\xi_c\propto 1/\kappa$. We then compute $t_{f,i}$ at finite $\kappa$ and extrapolate the result to the $\kappa \rightarrow 0$ limit, 
while keeping the distance between the two visons larger than $\xi_c$ throughout the extrapolation.
This procedure ensures that the second vison does not affect the local hopping process of the target vison. As discussed in the main text,
our calculation shows that $|t_{f,i}/J_3|$ extrapolates to zero in the $\kappa \rightarrow 0$ limit for both the FM and AFM Kitaev interactions.

A BVP is formed by two neighboring visons accompanied by a matter-fermion Bogoliubov excitation~\cite{Zhang_2021_vs,Chen_2025_ap}. 
It can be viewed as a particle residing on the bond separating the two visons. A BVP on the bond $(\br,\alpha)$, with $\br \in A$ and $\alpha = x,y,z$ 
(see \cref{fig:vison-move}(b)), is defined as
\begin{align}
    d_{\br,\alpha,l}^\dagger |\Omega \rangle = 2^{N-1/2} P\, \chi_{\br,\alpha}^\dagger \alpha_l^\dagger(\br,\alpha) |0; \Psi_c(\br,\alpha) \rangle.
\end{align}
Here $|\Omega \rangle$ denotes the ground state of the Kitaev model, and $\chi_{\br,\alpha}^\dagger = (b_{\br}^\alpha - i b_{\br+\bdelta_{\alpha}}^\alpha)/2$ 
is the creation operator for the gauge fermion on the bond $(\br,\alpha)$. The state $|0; \Psi_c(\br,\alpha) \rangle$ is the ground state of an 
effective matter-fermion BdG Hamiltonian $H_c(\br,\alpha)$ incorporating two neighboring $\pi$ fluxes, and $\alpha_{l}^\dagger(\br,\alpha)$ creates 
the $l$-th Bogoliubov quasiparticle of $H_c(\br,\alpha)$. The projection operator is $P \equiv \prod_j ( 1 + D_j )/2$, with $D_j = i b_j^x b_j^y b_j^z c_j$.
Here $N$ is the number of unit cells in the system, which is half of the number of sites.

% ------------------------------------------------------------------------------
% FIGURE
% ------------------------------------------------------------------------------
% ------------------------------------------------------------------------------

The effective tight-binding model for the BVPs has the form:
\begin{align}\label{eq:BVP-tb}
    H_\mathrm{b} = & \sum_{\br \in A,\alpha} d_{\br,\alpha}^\dagger \Delta_d d_{\br,\alpha} + 
    \sum_{\br \in A} \left[ d_{\br-\ba_1+\ba_2,x}^\dagger T_{-\ba_1+\ba_2}^{x} d_{\br,x} + d_{\br-\ba_1-\ba_2,x}^\dagger T_{-\ba_1-\ba_2}^{x} d_{\br,x} 
    \right. \nonumber \\
    & + d_{\br-\ba_2,y}^\dagger T_{-\ba_2}^y d_{\br,y} + d_{\br+2\ba_1-\ba_2,y}^\dagger T_{2\ba_1-\ba_2}^y d_{\br,y}
    \nonumber \\
    & \left. + d_{\br+\ba_1,z}^\dagger T_{\ba_1}^{z} d_{\br,z} + d_{\br-\ba_1+2\ba_2,z}^\dagger T_{-\ba_1 + 2\ba_2}^{z} d_{\br,z} + \mathrm{H.c.} \right].
\end{align}
Here $d_{\br,\alpha}^\dagger \equiv (d_{\br,\alpha,1}^\dagger, \dots, d_{\br,\alpha,N}^\dagger)$.
The diagonal matrix $\Delta_d$ contains the excitation energies of the BVPs. The hopping matrix $T_{\bR}^\alpha$ is generated by the $J_3$ interaction, 
and its matrix element is defined as
\begin{align}
    [T_{\bR}^\alpha]_{m,n} = \langle \Omega | d_{\br+\bR,\alpha,m}\, H_{J_3}\, d_{\br,\alpha,n}^\dagger |\Omega \rangle.
\end{align}
This matrix element can be evaluated using the Majorana representation of spins. Similar calculations for other non-Kitaev interactions 
can be found in Refs.~\cite{Joy_2022_do,Chen_2023_no,Chen_2025_ap,Chen_2025_apa}. Unlike nearest-neighbor non-Kitaev interactions such as the $\Gamma$ or Heisenberg terms, 
the $J_3$ interaction only induces hopping between BVPs of the same bond type. 
For example, \cref{fig:vison-move}(b) illustrates the hopping processes of $z$-bond BVPs: the hopping between the $d_{1,z,m}$ and $d_{5,z,n}$ modes 
is mediated by $\sigma_1^z \sigma_4^z$ and $\sigma_2^z \sigma_5^z$, while the hopping between the $d_{7,z,m}$ and $d_{3,z,n}$ modes is generated 
by $\sigma_3^z \sigma_6^z$. The hopping processes for $x$- and $y$-bond BVPs are related by a $C_3$ rotation.

We emphasize that the vanishing single-vison hopping obtained here applies only to the leading-order $J_3$-induced process. 
It does not exclude higher-order contributions, which may generate a weak single-vison dispersion at larger $J_3$, 
although such processes are expected to be subleading near the Kitaev limit. 
By contrast, the BVP sector acquires finite hopping already at leading order and therefore provides the dominant instability channel out of the KSL. 
The absence of leading single-vison hopping thus supports a regime in which magnetic order develops through the vison-pair channel while
single visons remain gapped. 
Higher-order single-vison processes may become relevant deeper in the ordered regime,  possibly contributing to the eventual confinement
into conventional magnetic phases.

\subsection{H. Magnetic order from the condensation of BVPs}
After obtaining the hopping matrices $T^\alpha_{\bR}$, we rewrite $H_\mathrm{b}$ in momentum space by Fourier transforming the BVP modes,
$d_{\bq,\alpha,l}^\dagger = \frac{1}{\sqrt{N}} \sum_{\br \in A} e^{i \bq \cdot \br} d_{\br,\alpha,l}^\dagger$.
Diagonalizing $H_\mathrm{b}$ then gives the BVP quasiparticle modes. The lowest-energy mode at momentum $\bq$ can be written as
\begin{align}
    \beta_{\bq}^\dagger = d_{\bq,x}^\dagger w_x(\bq) + d_{\bq,y}^\dagger w_y(\bq) + d_{\bq,z}^\dagger w_z(\bq),
\end{align}
where $w_{\alpha}(\bq)$ is an $N \times 1$ vector describing the weight of this mode on the $d_{\bq,\alpha}$ BVP states.
Here we have used the shorthand notation $d_{\bq,\alpha}^\dagger = ( d_{\bq,\alpha,1}^\dagger, \dots, d_{\bq,\alpha,N}^\dagger )$.

When the $\beta_{\bq}$ mode condenses, the ground state can be written perturbatively as
\begin{align}
    |\Psi \rangle =  \left( 1 + \lambda e^{i \theta} \beta_{\bq}^\dagger + \dots \right) |\Omega \rangle,
\end{align}
where $\lambda$ is the condensate amplitude and $\theta$ is the condensate phase. To leading order in $\lambda$, the induced local moment is given by
\begin{align}\label{eq:BVP_moments}
    \langle \sigma_{\br}^\alpha \rangle \propto e^{i \theta} \langle \Omega | \sigma_{\br}^\alpha \beta_{\bq}^\dagger | \Omega \rangle + \mathrm{H.c.}
\end{align}
For $\br \in A$, one obtains
\begin{align}\label{eq:BVP_moments_Asub}
     \langle \Omega | \sigma_{\br}^\alpha \beta_{\bq}^\dagger | \Omega \rangle 
     = & \frac{1}{\sqrt{N}} e^{i \bq \cdot \br} \langle \Omega | \sigma_{\br}^\alpha d_{\br,\alpha}^\dagger | \Omega \rangle w_{\alpha}(\bq) \nonumber \\
     = & \frac{1}{\sqrt{N}} e^{i \bq \cdot \br} P_{\alpha,A} w_\alpha(\bq),
\end{align}
where we have defined $P_{\alpha,A} \equiv \langle \Omega | \sigma_{\br}^\alpha ( d_{\br,\alpha,1}^\dagger, \dots, d_{\br,\alpha,N}^\dagger )|\Omega \rangle$.
Similarly, for $\br' \in B$, one finds
\begin{align}\label{eq:BVP_moments_Bsub}
    \langle \Omega | \sigma_{\br'}^\alpha \beta_{\bq}^\dagger | \Omega \rangle
    = & \frac{1}{\sqrt{N}} e^{i \bq \cdot (\br' - \bdelta_\alpha)} \langle \Omega | \sigma_{\br'}^\alpha 
    d_{\br'-\bdelta_\alpha,\alpha}^\dagger |\Omega \rangle \nonumber \\
    = & \frac{1}{\sqrt{N}} e^{i \bq \cdot (\br' - \bdelta_\alpha)} P_{\alpha,B} w_{\alpha}(\bq),
\end{align}
with $P_{\alpha,B} \equiv \langle \Omega | \sigma_{\br'}^\alpha ( d_{\br'-\bdelta_\alpha,\alpha,1}^\dagger, \dots, d_{\br'-\bdelta_\alpha,\alpha,N}^\dagger )|\Omega \rangle$.
Therefore, once the eigenvector $w_\alpha(\bq)$ and the matrix elements $P_{\alpha,A/B}$ are computed, the magnetic ordering pattern induced by the condensation of $\beta_{\bq}$ follows directly from \cref{eq:BVP_moments,eq:BVP_moments_Asub,eq:BVP_moments_Bsub}.
For both the $\beta_{\Gamma}$-induced AFM order and the $\beta_{\M_i}$-induced ZZ$_{1,2}$ order, the resulting ordering pattern and spin-axis direction are 
independent of the condensate phase $\theta$, up to symmetry-related choices such as translations or time reversal, similar to the case of other NN non-Kitaev interactions~\cite{Chen_2025_ap,Zhang_2021_vs}.

\end{document}